\documentclass[%
 reprint,
 amsmath,amssymb,
 aps,
]{revtex4-2}

\usepackage{graphicx}
\usepackage{dcolumn}
\usepackage{bm}
\usepackage{xspace}
\usepackage{xcolor}

\newcommand{\PbPb}         {\mbox{Pb--Pb}\xspace}
\newcommand{\XeXe}         {\mbox{Xe--Xe}\xspace}
\newcommand{\UU}         {\mbox{U--U}\xspace}
\newcommand{\NeNe}         {\mbox{Ne--Ne}\xspace}
\newcommand{\Pb}         {\mbox{$^{208}$Pb}\xspace}
\newcommand{\Xe}         {\mbox{$^{129}$Xe}\xspace}

\newcommand{\Ne}         {\mbox{$^{20}$Ne}\xspace}
\newcommand{ \la }{\langle}
\newcommand{ \ra }{\rangle}

\newcommand{\nbodynleft}{$8654$\xspace}
\newcommand{\ntipnleft}{$40773$\xspace}
\newcommand{\nbodyfracnleft}{$1.84\%$\xspace}
\newcommand{\ntipfracnleft}{$8.67\%$\xspace}

\newcommand{\nbodyfracnpgcm}{$1.78\%$\xspace}
\newcommand{\ntipfracnpgcm}{$7.60\%$\xspace}

\newcommand{\nlefttippurity}{$0.573$\xspace}
\newcommand{\nleftbodypurity}{$0.201$\xspace}
\newcommand{\PGCMtippurity}{$0.540$\xspace}
\newcommand{\PGCMbodypurity}{$0.156$\xspace}

\begin{document}

\preprint{APS/123-QED}

\title{Nuclear-geometry-tagging events in high-energy \NeNe collisions}

\author{Joachim Carlo Kristian Bjerg Hansen}
\email{joachim.hansen@fysik.lu.se}
 \affiliation{Lund University}
\author{Emil Gorm Dahlbæk Nielsen}%
 \email{emil.gorm.nielsen@cern.ch}
\affiliation{CERN
}%

\date{\today}

\begin{abstract}

Nucleon-nucleon interactions within atomic nuclei give rise to collective properties such as deformation and nuclear clustering. Ultrarelativistic nucleon-nucleon collisions offer a novel way to probe the many-body correlations of the incoming nuclei, as the nuclear structure is reflected in the initial geometry of the produced system. State-of-the-art low-energy nuclear theory calculations predict highly deformed $^{20}$Ne nuclei with a characteristic $\alpha$+$^{16}$O bowling-pin-like shape. Since such nuclei have recently been collided at the LHC, investigating methods to maximise the sensitivity to the nuclear structure is pertinent. In this paper, we investigate whether body--body and tip--tip configurations of $^{20}$Ne nuclei can be identified in Monte--Carlo Glauber simulations using geometry-sensitive observables: the eccentricity, $\varepsilon_2$, and the transverse compactness, $d_\perp$. The nuclear configurations are obtained from nuclear lattice effective field theory calculations within the framework of Nuclear Lattice Effective Field Theory (NLEFT) and \textit{ab initio} Projected Generator Coordinate Method (PGCM). We utilise a Fisher linear discriminant to select body--body and tip--tip configurations based on the $\varepsilon_2$-$d_\perp$ correlation. With classifier thresholds chosen to retain $30\%$ of the target class in a validation sample, the two-variable Fisher selection obtains purities in NLEFT (PGCM) of \nleftbodypurity (\PGCMbodypurity) and \nlefttippurity (\PGCMtippurity) for body--body and tip--tip events, respectively.

\end{abstract}

\maketitle


\section{Introduction}
One of the key signatures of the formation of a strongly interacting quark--gluon plasma (QGP) is the observation of a large anisotropy in the azimuthal distribution of produced particles~\cite{Teaney:2000cw,STAR:2005gfr,PHENIX:2004vcz,PHOBOS:2004zne,ALICE:2022wpn, CMS:2024krd}. This anisotropy arises from the fluid-like evolution of the QGP, in which pressure gradients convert the initial spatial anisotropy into final-state momentum anisotropy in the transverse (x, y) plane perpendicular to the collision axis. The resulting anisotropy is commonly characterized by a Fourier expansion of the azimuthal particle distribution~\cite{Voloshin:1994mz}
\begin{align}
\frac{dN}{d\varphi}\propto 1+2\sum_{n=1}^\infty v_n\cos \left[n(\varphi-\Psi_n)\right]
\end{align}
where $\varphi$ is the azimuthal angle, $v_n$ is the flow coefficient and $\Psi_n$ is the symmetry plane angle. In general, spatial anisotropy is generated by the geometric overlap of the matter densities of the colliding nuclei, leading to the characteristic centrality dependence of $v_2$ observed in heavy-ion collisions. The $v_2$ is, to leading order and within the hydrodynamic paradigm, proportional to the initial eccentricity $\varepsilon_2$~\cite{Gardim:2011xv,Niemi:2012aj,Noronha-Hostler:2015dbi}. For non-spherical nuclei in head-on collisions, the initial geometry of the system can also reflect the intrinsic nuclear deformation, thereby influencing the subsequent QGP evolution even in the absence of a finite impact parameter~\cite{STAR:2024wgy,Jia:2025wey,Giacalone:2026fat,Giacalone:2023hwk,Blaizot:2025scr,Blaizot:2025bfu}. 
The effect of nuclear deformation on anisotropic flow observables has been observed at both RHIC~\cite{STAR:2024wgy} and the LHC~\cite{ALICE:2018lao,ATLAS:2019dct,CMS:2019cyz}. Another observable that characterises the initial geometry is the mean transverse momentum, $\la [p_\mathrm{T}]\ra$, of charged particles in the event. At fixed multiplicity, relative fluctuations of the mean transverse momentum, $\delta\la [p_\mathrm{T}]\ra/\la [p_\mathrm{T}]\ra$, are expected to be proportional to fluctuations in the initial inverse transverse size of the system $\delta d_\perp/\la d_\perp\ra$: a smaller initial transverse area corresponds to a larger $d_\perp$, a higher entropy density, and consequently stronger radial flow~\cite{Giacalone:2020dln,Jia:2021qyu}. The correlation between elliptic flow and mean transverse momentum has proven a powerful constraint on nuclear geometry and initial-state structure~\cite{Bozek:2016yoj,Giacalone:2020awm,Giacalone:2021clp,Jia:2021qyu,Dimri:2023wup,Zhao:2024lpc,Nijs:2022rme}, and has also been measured experimentally at the LHC for \XeXe and \PbPb collisions~\cite{ALICE:2021gxt,ATLAS:2022dov}. Recent studies have even used Bayesian parameter estimation techniques to constrain low-energy nuclear structure parameters for \Xe and \Pb~\cite{Giacalone:2026fat}.

In most theoretical modelling, nuclear deformation is incorporated through modified Woods–Saxon distributions, which effectively treat nuclei as one-body nuclear densities. More recently, the use of state-of-the-art many-body nuclear-structure inputs has been successful in describing collisions of $^{16}$O and $^{20}$Ne at the LHC~\cite{ALICE:2025luc}. This suggests that the ultrarelativistic collisions of deformed nuclei can be used as a probe of the many-body interactions of nuclear matter and open up an interesting new way to test predictions from realistic many-body, low-energy nuclear theory calculations. In particular, $^{20}$Ne is expected to possess one of the most extreme ground-state geometries. Low-energy theoretical calculations
using Nuclear Lattice Effective Field Theory (NLEFT) or Projected Generator Coordinate Method (PGCM) predict that $^{20}$Ne is highly deformed with a characteristic $\alpha$+$^{16}$O bowling-pin-like shape~\cite{Giacalone:2024luz}. 

The effects of nuclear deformation are generally expected to be most pronounced in central collisions, whereas in non-central collisions they compete with anisotropies arising from the geometric overlap of the colliding nuclei. Specific configurations in which the principal axes of the nuclei are aligned, such as body--body and tip--tip collisions, maximise the sensitivity of the initial geometry to the nuclear structure. In the most central collisions, approximately $2A$ nucleons participate; however, because the nuclei are randomly oriented, these specific configurations constitute only a small fraction of the central event sample. A method for separating them from other configurations is therefore needed to obtain the maximal sensitivity to nuclear-structure effects. Such a separation has previously been performed in \UU collisions~\cite{Goldschmidt:2015kpa}, where tip--tip and body--body events were selected using multiplicity, spectator number, and \(v_2\). It remains to be established whether including the inverse transverse size improves such geometry tagging and whether the method is effective for the asymmetric many-body structure of \Ne.

\begin{figure}[t!]
  \centering
  \includegraphics[width=\linewidth]{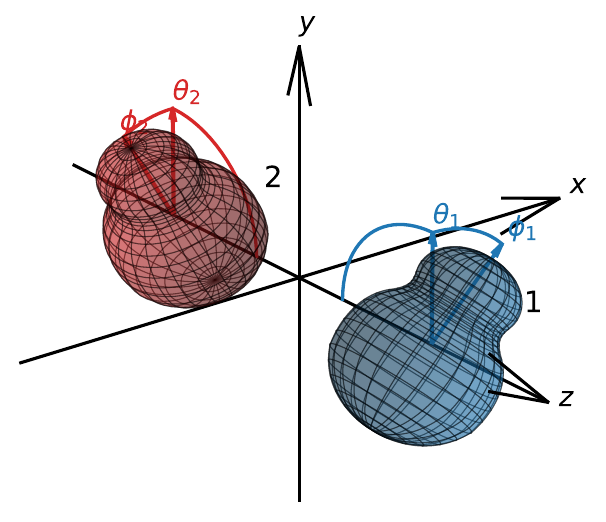}
  \caption{Schematic illustration of two pin–shape nuclei and the angles used to
  describe their orientation relative to the beam axis.}
  \label{fig:axis_definition}
\end{figure}
In this paper, we employ a linear separator to separate body--body and tip--tip \NeNe collisions from the general event ensemble based on the correlation between second-order eccentricity, $\varepsilon_2$, and the relative fluctuations of the inverse transverse size, $\delta d_\perp/\la d_\perp\ra$, to enhance the sensitivity to the geometry of $^{20}$Ne. The two observables are provide complementary sensitivity to the shape and inverse transverse size of the system and are related to the final-state observables $v_2$ and $\la [p_\mathrm{T}]\ra$. We study the shape and size of $^{20}$Ne by analysing the eccentricity, $\varepsilon_2$, and the inverse transverse size, $\delta d_\perp/\la d_\perp\ra$ in body--body and tip--tip subsets of the randomly oriented ensemble of the colliding nuclei within the Monte--Carlo Glauber framework. 

The paper is organized in the following way. In Section \ref{sec:method}, the MC Glauber model and the geometry-sensitive observables, $\varepsilon_2$ and $\delta d_\perp/\la d_\perp\ra$, are discussed. In section \ref{sec:separators}, we discuss the implementation of the linear separation. Finally, the results are summarised in Section \ref{sec:conclusion}.

\section{Methods}
\label{sec:method}
In this section, the simulation setup and the observables used to characterize the
collision geometry are introduced. First, the nuclear configurations and the
Monte--Carlo Glauber model are described. The orientation classes and the
geometry-sensitive observables used throughout the analysis are then defined.

\subsection{Collision model}

\begin{figure}[t!]
  \centering
  \includegraphics[width=\linewidth]{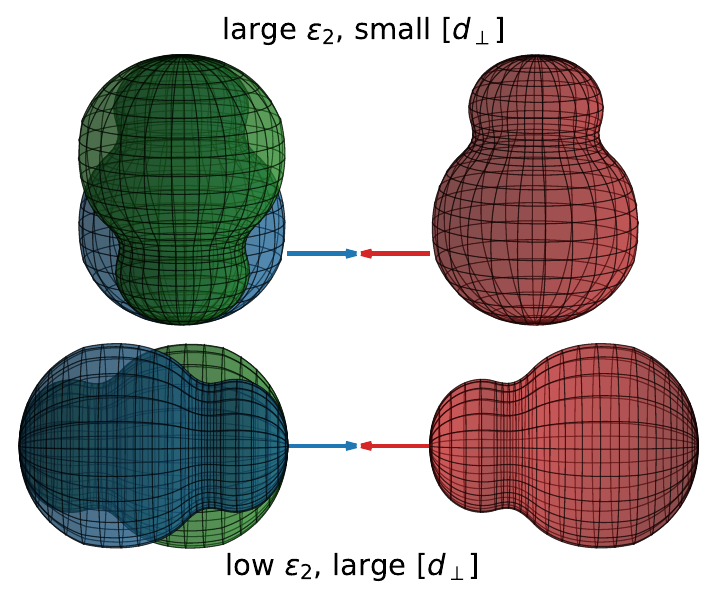}
  \caption{Schematic illustration of the different configurations that correspond to \textit{pure} tip–tip and body–body.}
  \label{fig:axis_configurations}
\end{figure}
The initial geometry of the Ne--Ne collisions is simulated using a Monte--Carlo
Glauber model. As a spherical reference, the nucleon positions are sampled from a
Fermi distribution of the form
\begin{align}
  \rho(r) = \rho_0 \frac{1 + w \frac{r^2}{R^2}}{1 + e^{\frac{r - R}{a}}},
\end{align}
where $\rho_0$ is the central density, $R$ is the half-density radius, $a$ is the
surface diffuseness, and $w$ describes deviations from the standard Woods--Saxon
shape.

For the deformed $^{20}$Ne nuclei, the spatial distribution of the nucleons is
instead obtained from nuclear lattice effective field theory (NLEFT) configurations
or the projected generator coordinate method (PGCM)~\cite{PGCM}.
These many-body configurations provide a more realistic description of the nuclear
structure than a one-body density and have been shown to reproduce important
features of recent light-ion collision measurements at the LHC~\cite{Giacalone:2024luz}. 

The nucleon--nucleon inelastic cross section is set to
$\sigma_{\mathrm{NN}}=68.0$~mb, corresponding to Ne--Ne collisions at
$\sqrt{s_{\mathrm{NN}}}=5.36$~TeV~\cite{MCGlauberNew}. A probabilistic
nucleon--nucleon overlap profile with $\omega=0.3$ is used, and a minimum
intra-nuclear nucleon distance of $0.4$~fm is imposed. Thirty-five million
events are generated for both the NLEFT and PGCM configurations. The NLEFT sample consists of 20 000 configurations, which are randomly selected per event, after which a random 3D rotation is applied. The PGCM sample instead samples the nucleon positions from the proton and neutron density grids each event, and then applies the random 3D rotation.

For the orientation-selection analysis to mimic experimental cuts, events are selected with $N_\mathrm{spec}\leq 2$, which roughly corresponds to the $1\%$ of events with the lowest total
spectator count, this represent the most central head–on collisions. Unless otherwise stated, the results are obtained from this central event sample. 

\begin{figure}[!t]
  \centering
  \includegraphics[scale=0.45]{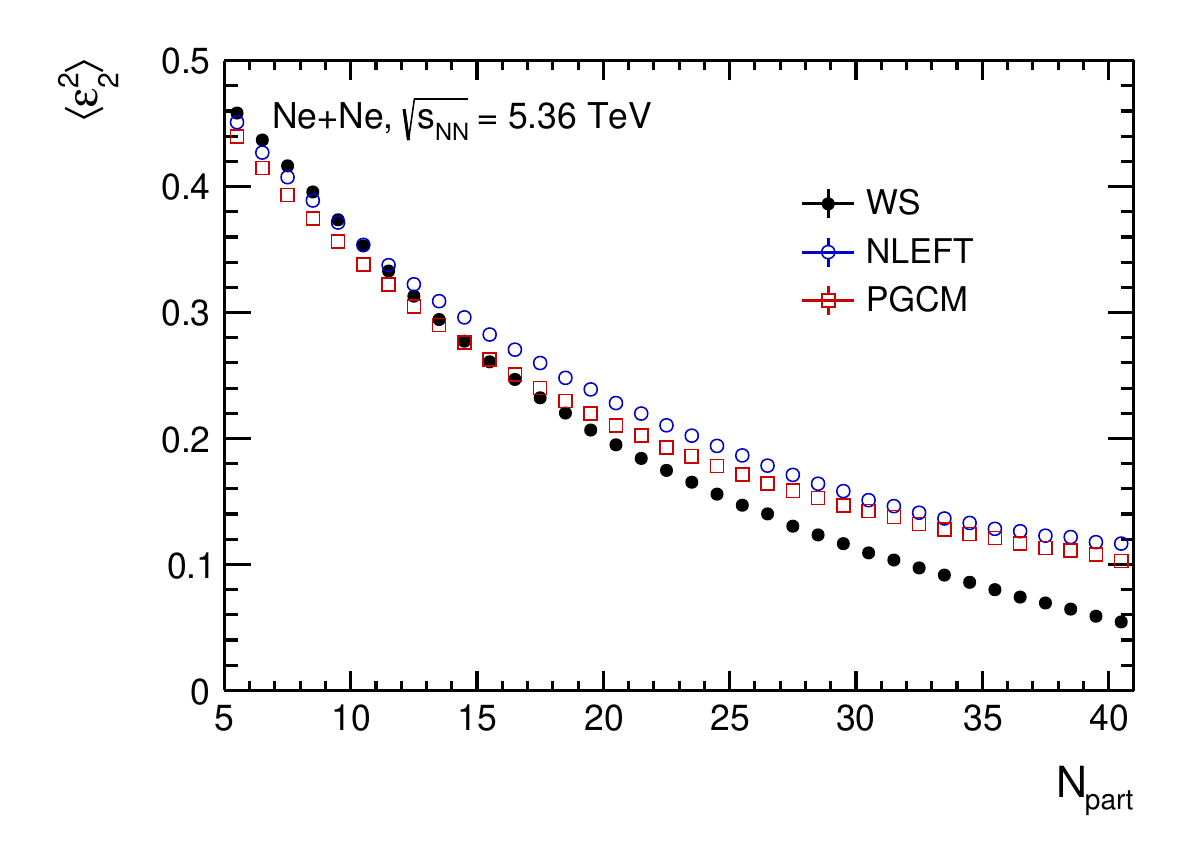}
  \caption{$\langle\varepsilon_2^2\rangle$ as a function of
  $N_\mathrm{part}$ for spherical and deformed $^{20}$Ne configurations}
  \label{fig:spherical_nleft_comp}
\end{figure}

\subsection{Orientation classes}
The orientation of each nucleus is described by the polar angle $\theta_i$ between
its major symmetry axis and the beam direction and by the azimuthal angle $\phi_i$
in the transverse plane. The index $i=1,2$ denotes the two incoming nuclei, as
illustrated in Fig.~\ref{fig:axis_definition}. Events are classified as tip--tip
when
\[
  \sqrt{\cos^2\theta_1 + \cos^2\theta_2} > 1.22,
\]
which selects events where both major axes are approximately aligned with the beam. The polar constraint implies that, for $\theta_1 \approx \theta_2$, tip--tip events have tilts of at most approximately $\pi/6$ relative to the beam direction.
The body--body class is defined by
\[
  \sqrt{\cos^2\theta_1 + \cos^2\theta_2} < 0.31,
\]
which is a polar constraint of around $\pi/10$, combined with the requirement
\[
  \Delta\phi_\mathrm{axis}
  =
  \min\left(|\Delta\phi|,\pi-|\Delta\phi|\right)
  < \frac{\pi}{10},
\]
where $\Delta\phi$ is wrapped to the interval $[-\pi,\pi]$. This requirement
accounts for the fact that the nuclear major axis is unchanged by a rotation of
$\pi$. All events that do not satisfy the target orientation criteria are referred
to as the rest of the event sample. The central sample for NLEFT contains \nbodynleft body--body events and \ntipnleft tip--tip events, corresponding to selected
fractions of \nbodyfracnleft and \ntipfracnleft, respectively. For PGCM the body--body, and tip--tip selected fractions are \nbodyfracnpgcm and \ntipfracnpgcm, respectively. The selection criteria were systematically varied to assess their influence on the event classification. Modifying the $\sqrt{\cos^2\theta_1 + \cos^2\theta_2}$ requirement by up to $10\%$ primarily affects the fraction of selected tip–tip events, while the body–body fraction changes only slightly. Similarly, relaxing the $\Delta\phi_\mathrm{axis}$ criterion to $\pi/5$ produces a comparable effect to varying the $\theta_n$ requirement.

\begin{figure}[t!]
  \centering
  \includegraphics[scale=0.45]{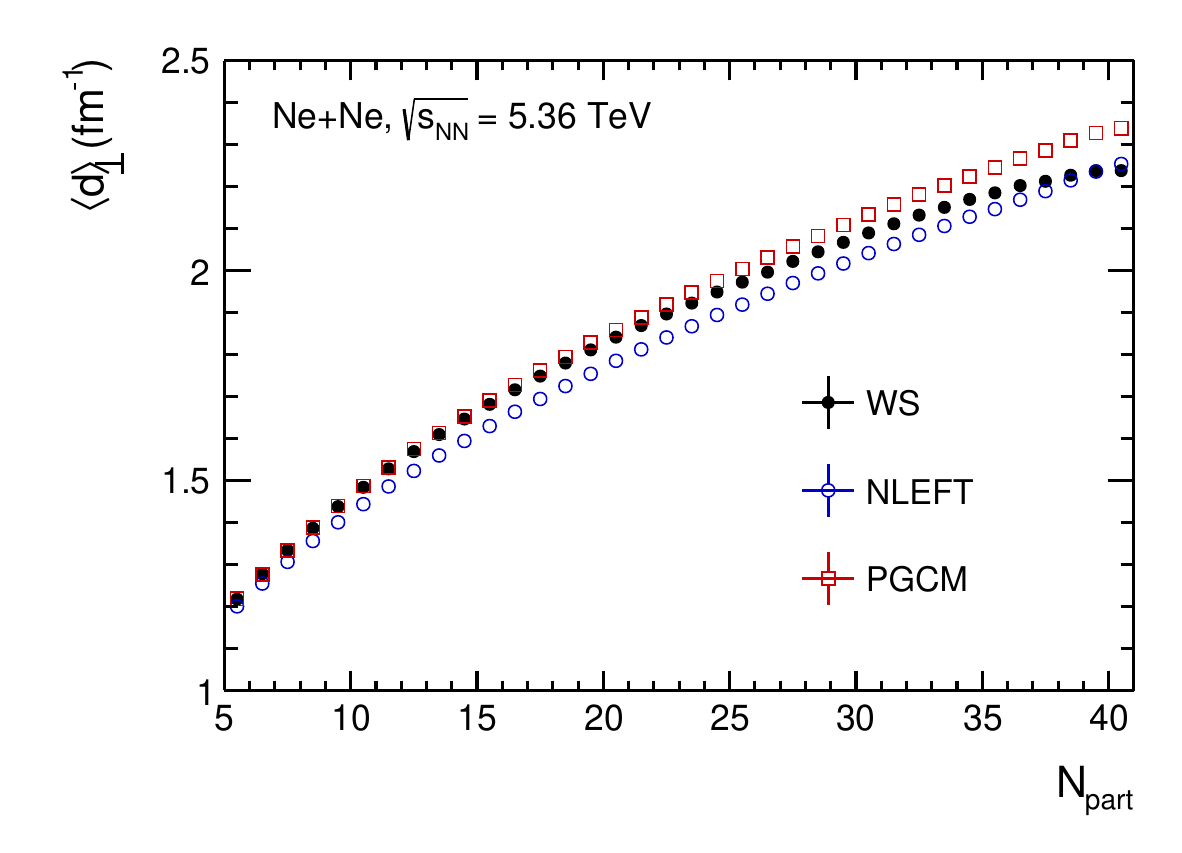}
  \caption{$\langle d_\perp \rangle $ as a function of
  $N_\mathrm{part}$ for spherical and deformed $^{20}$Ne configurations}
  \label{fig:spherical_nleft_comp_dperp}
\end{figure}

\begin{figure*}[t!]
  \centering
  \includegraphics[scale=0.4]{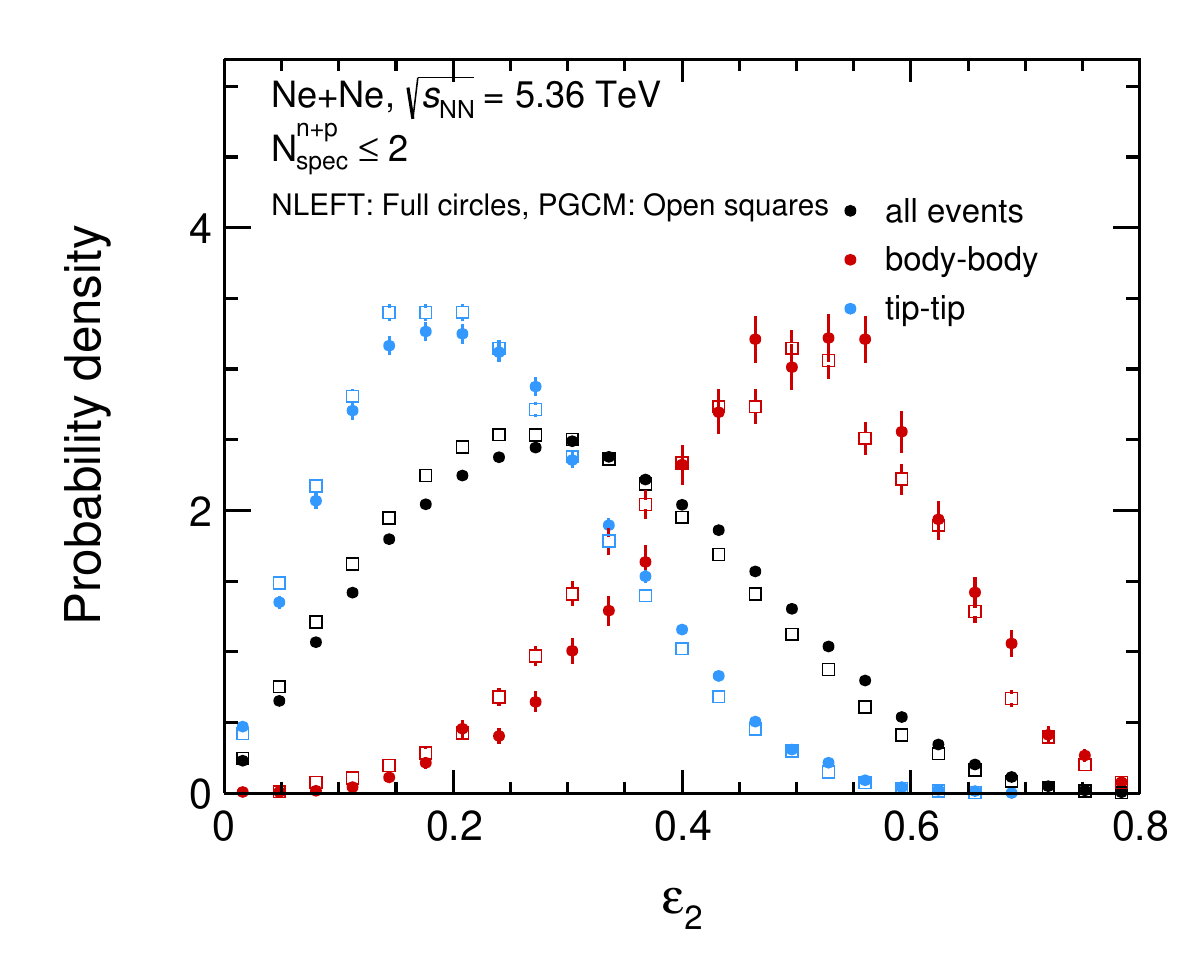}
  \includegraphics[scale=0.4]{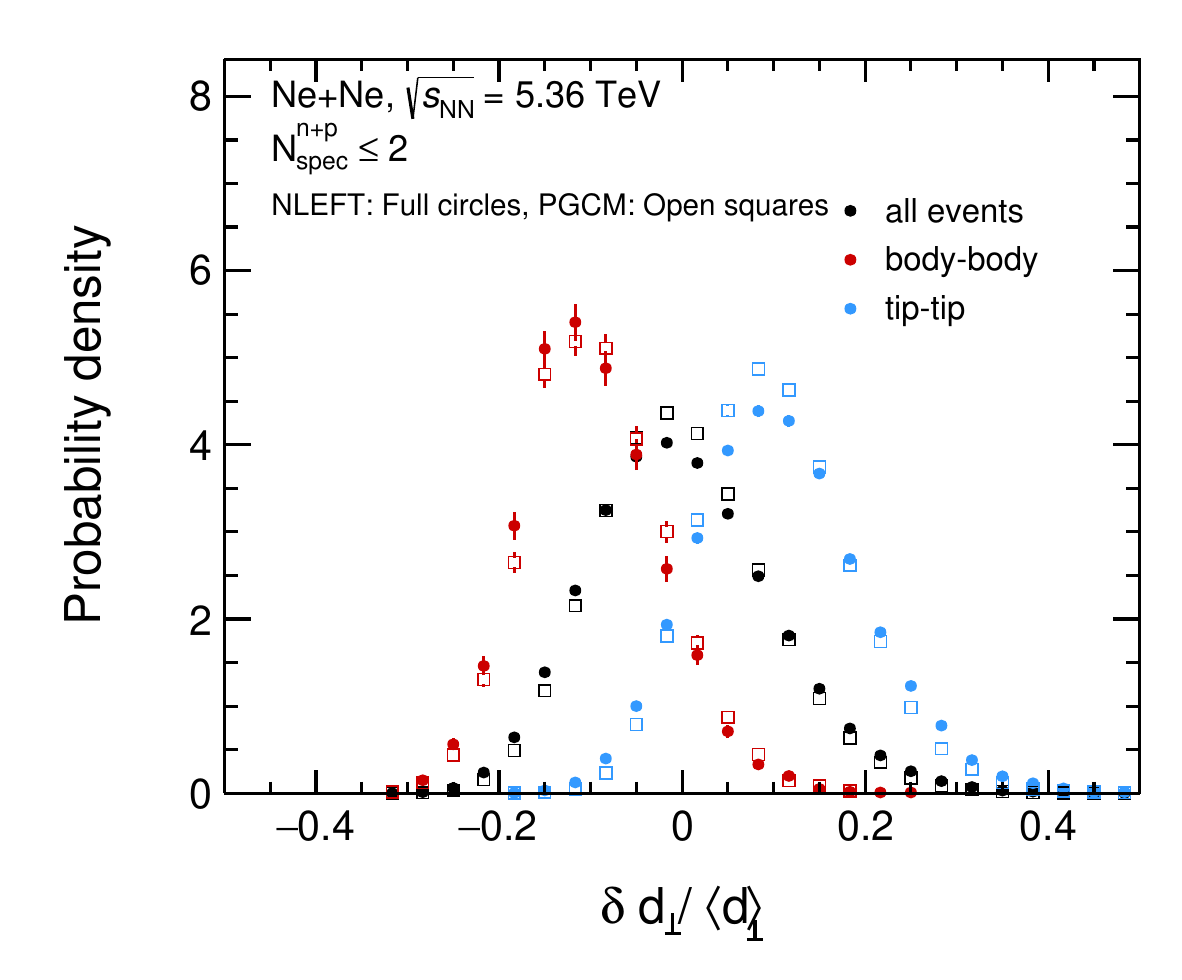}
  \caption{Probability density distributions of $\varepsilon_2$ (left) and
  $[d_\perp]$ (right) for inclusive, body--body, and tip--tip Ne--Ne collisions
  with $N_\mathrm{spec} \leq 2$. NLEFT is represented by the full circles and PGCM by the open squares.}
  \label{fig:P_ecc2_dperp}
\end{figure*}

\subsection{Geometry-sensitive observables}
The eccentricities are calculated relative to the centre of the participant
distribution as
\begin{align}
  \varepsilon_n
  =
  \frac{\left|\sum_j r_j^n e^{in\phi_j}\right|}
  {\sum_j r_j^n},
\end{align}
where the sum runs over all participant nucleons. The first observable used in the
orientation analysis is the complex ellipticity,
\begin{align}
  \epsilon_2 = \varepsilon_2 e^{i2\Phi_2},
\end{align}
where $\varepsilon_2$ is the magnitude of the elliptic geometry and $\Phi_2$ is
the corresponding participant-plane angle. Both quantities are calculated from
the transverse positions of the participant nucleons.

The second observable describes the compactness of the participant distribution.
It is constructed from the transverse overlap area~\cite{Jia:2021qyu} and the number of participants, $N_\mathrm{part}$ as 
\begin{align}
  S_\perp = \pi \sqrt{\langle x^2 \rangle \langle y^2 \rangle},
\end{align}
where $\langle x \rangle$ and $\langle y \rangle$ denote the transverse spatial coordinates of the participating nucleons, followed by
\begin{align}
  d_\perp = \sqrt{\frac{N_\mathrm{part}}{S_\perp}}.
\end{align}
To compare events with different absolute compactness, the normalised fluctuation
is defined as
\begin{align}
  [d_\perp]
  =
  \frac{\delta d_\perp}{\langle d_\perp\rangle},
  \qquad
  \delta d_\perp = d_\perp-\langle d_\perp\rangle.
\end{align}
The mean $\langle d_\perp\rangle$ is evaluated separately within each integer
$N_\mathrm{part}$ interval. This removes the leading dependence on collision
centrality before the orientation selections are constructed.

The two observables contain complementary information, see Fig.~\ref{fig:axis_configurations}. A body--body collision is expected to produce a more elliptic participant distribution and therefore a larger $\varepsilon_2$. A tip--tip collision has a smaller transverse area and is expected
to produce a larger $d_\perp$. Event-by-event fluctuations broaden both distributions and prevent a perfect separation of the orientation classes.
Experimentally, $\varepsilon_2$ and $d_\perp$ are not measured directly, but they are related to final-state observables such as elliptic flow and the mean transverse momentum. 
Selections in these quantities can therefore serve as a model-level description of a possible event-shape engineering analysis.



\begin{figure*}[t!]
  \centering
  \includegraphics[scale=0.4]{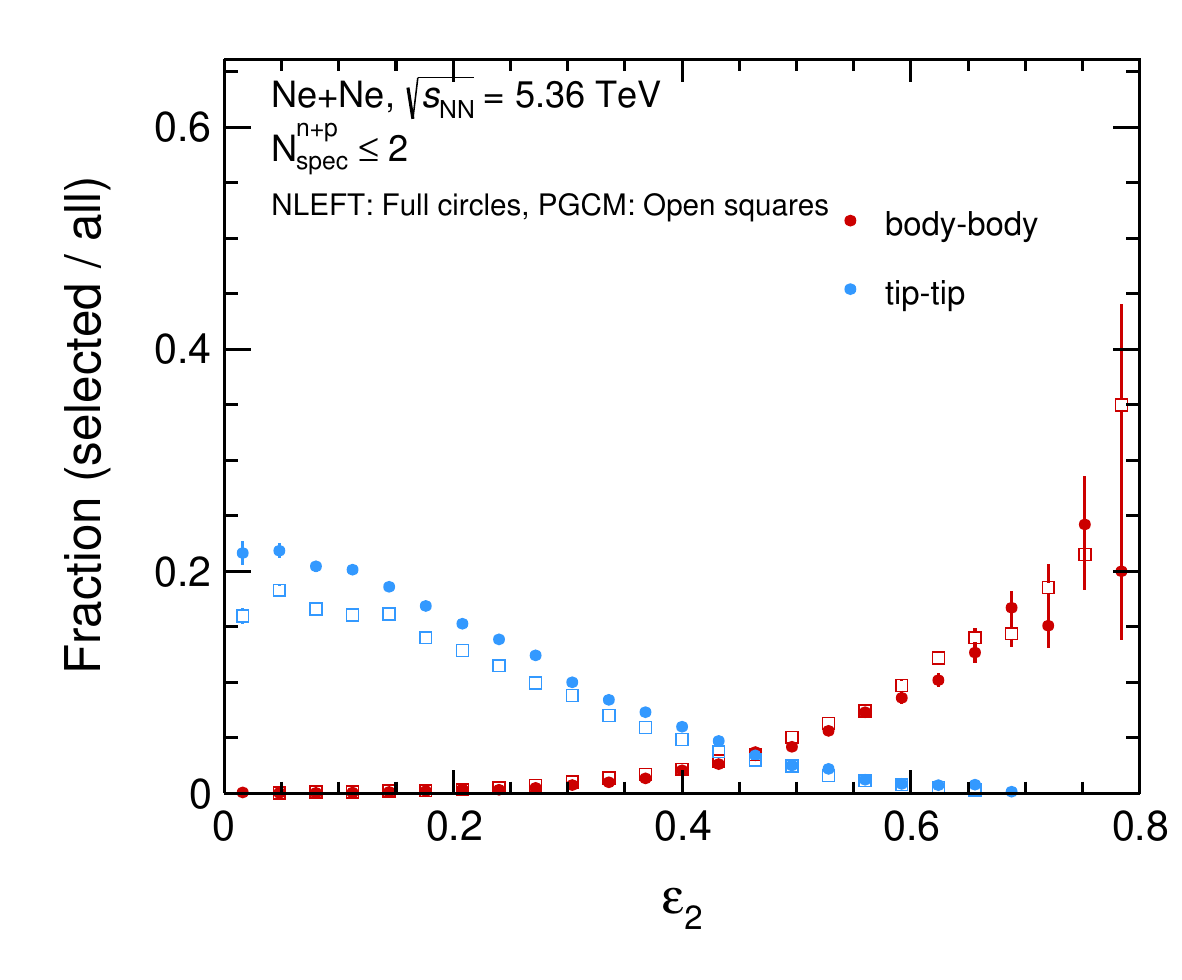}
  \includegraphics[scale=0.4]{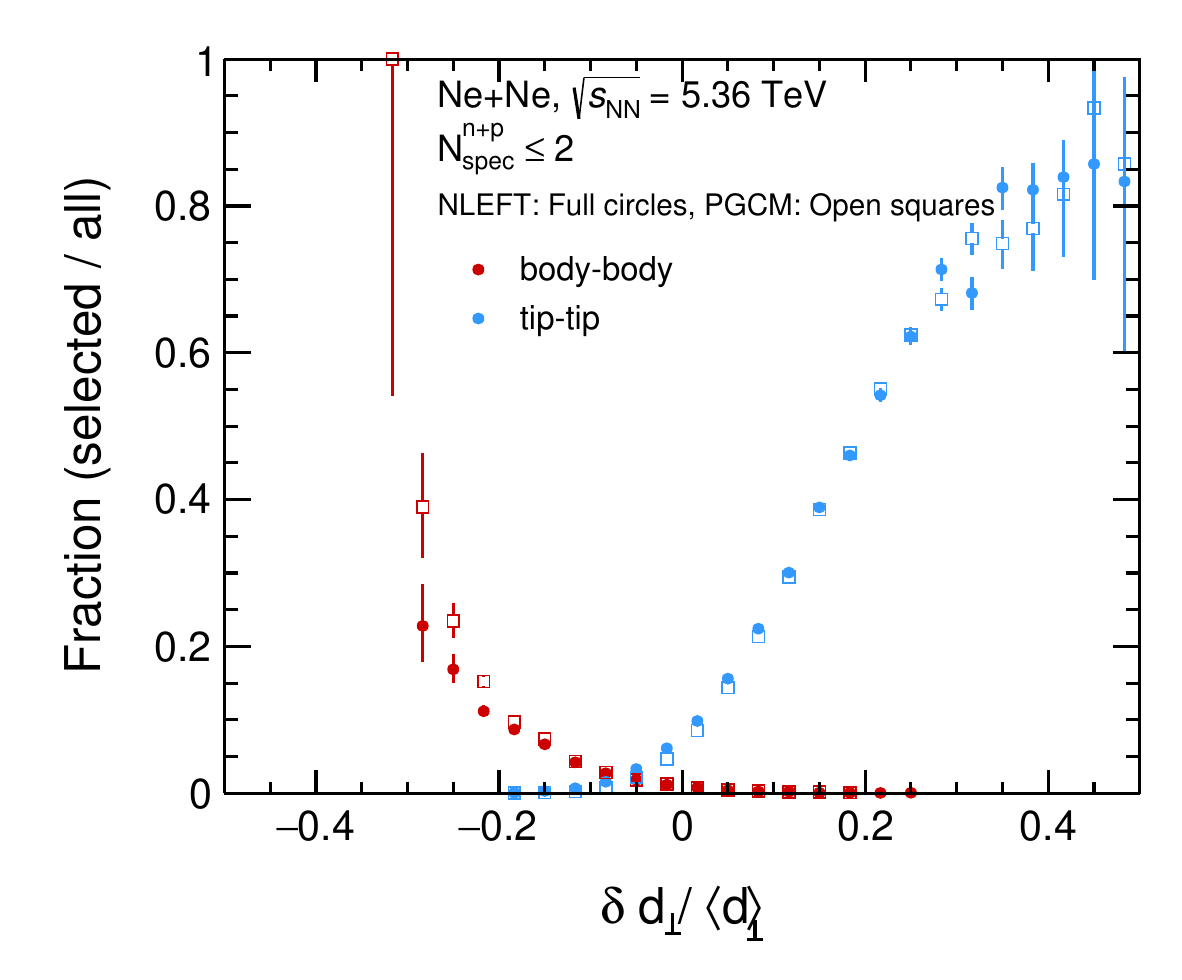}
  \caption{Fractions of body--body and tip--tip events as a function of
  $\varepsilon_2$ (left) and $[d_\perp]$ (right) for
  $N_\mathrm{spec} \leq 2$. NLEFT is represented by the full circles and PGCM by the open squares.}
  \label{fig:eff_ecc2_dperp}
\end{figure*}

\section{Sensitivity to deformation beyond central collisions}
The effect of the nuclear structure on $\varepsilon_2^2$ is studied as a function of $N_\mathrm{part}$ in Figure.~\ref{fig:spherical_nleft_comp}. The figure compares $\langle\varepsilon_2^2\rangle$ from spherical Fermi configurations with the deformed NLEFT and PGCM configurations. The largest difference between the spherical and deformed configurations is
observed at high $N_\mathrm{part}$, where the intrinsic nuclear shape contributes most directly to the collision geometry. However, the difference remains visible down to approximately $N_\mathrm{part}\sim13$, which show that $\varepsilon_2$ can retain sensitivity to the nuclear structure over a wider range of $N_\mathrm{part}$, and therefore, centrality. The figure also shows that at high $N_\mathrm{part}$ there are little differences between NLEFT and PGCM, but as the number of participants decrease, considerable differences appear between $13 < N_\mathrm{part} < 25$, providing a window for probing their differences in experiments. Fig.~\ref{fig:spherical_nleft_comp_dperp} presents the $\langle d_\perp \rangle$ as a function of $N_\mathrm{part}$ for the three different geometrical configurations. In the most central collisions at large $N_\mathrm{part}$, the biggest sensitivity to differences in $d_\perp$ appears to be in PGCM, whereas NLEFT show differences at lower multiplicity.

The observed differences in $d_\perp$ between the nuclear structure models become more pronounced at high multiplicities, particularly between the NLEFT and PGCM predictions. This may indicate a sensitivity of the transverse compactness to the underlying nuclear structure, reflecting differences in the spatial distributions of participating nucleons. Unlike the eccentricity, which characterizes the anisotropy of the initial geometry, $d_\perp$ provides complementary information about the transverse size and density of the collision system. Such differences could potentially influence final-state observables, such as the mean transverse momentum.

\section{Single-observable selections}
Figure~\ref{fig:P_ecc2_dperp} presents the probability density distributions of
$\varepsilon_2$ and $[d_\perp]$ for events with $N_\mathrm{spec} \leq 2$. The
body--body distribution is shifted towards larger $\varepsilon_2$, as expected
from the elongated transverse overlap of the two nuclei. In contrast, the
tip--tip distribution is shifted towards larger $[d_\perp]$, which reflects the
smaller and more compact transverse overlap area. The distributions overlap considerably with the inclusive event sample. A
selection in a single observable can therefore enrich an orientation class, but it
cannot isolate a pure body--body or tip--tip sample. This overlap is a direct
consequence of the fluctuations in the nucleon positions and of the remaining
variation within each orientation class.

The corresponding orientation fractions are shown in
Fig.~\ref{fig:eff_ecc2_dperp}. The body--body contribution increases with
$\varepsilon_2$, while the tip--tip contribution is strongly suppressed in the
same region. At small $\varepsilon_2$, the opposite behaviour is observed, but the
tip--tip enhancement remains limited because low eccentricity is also produced by
many events without a specific orientation. The selection in $[d_\perp]$ is particularly effective for tip--tip events. The selection efficiency of tip--tip increases rapidly towards large positive values of $[d_\perp]$,
whereas the body--body fraction remains comparatively small. These results show
that $\varepsilon_2$ and $[d_\perp]$ respond differently to the nuclear orientation and motivate a combined selection.

\section{Combined event-shape selection}\label{sec:results}

\begin{figure*}[t!]
  \centering
  \includegraphics[scale=0.4]{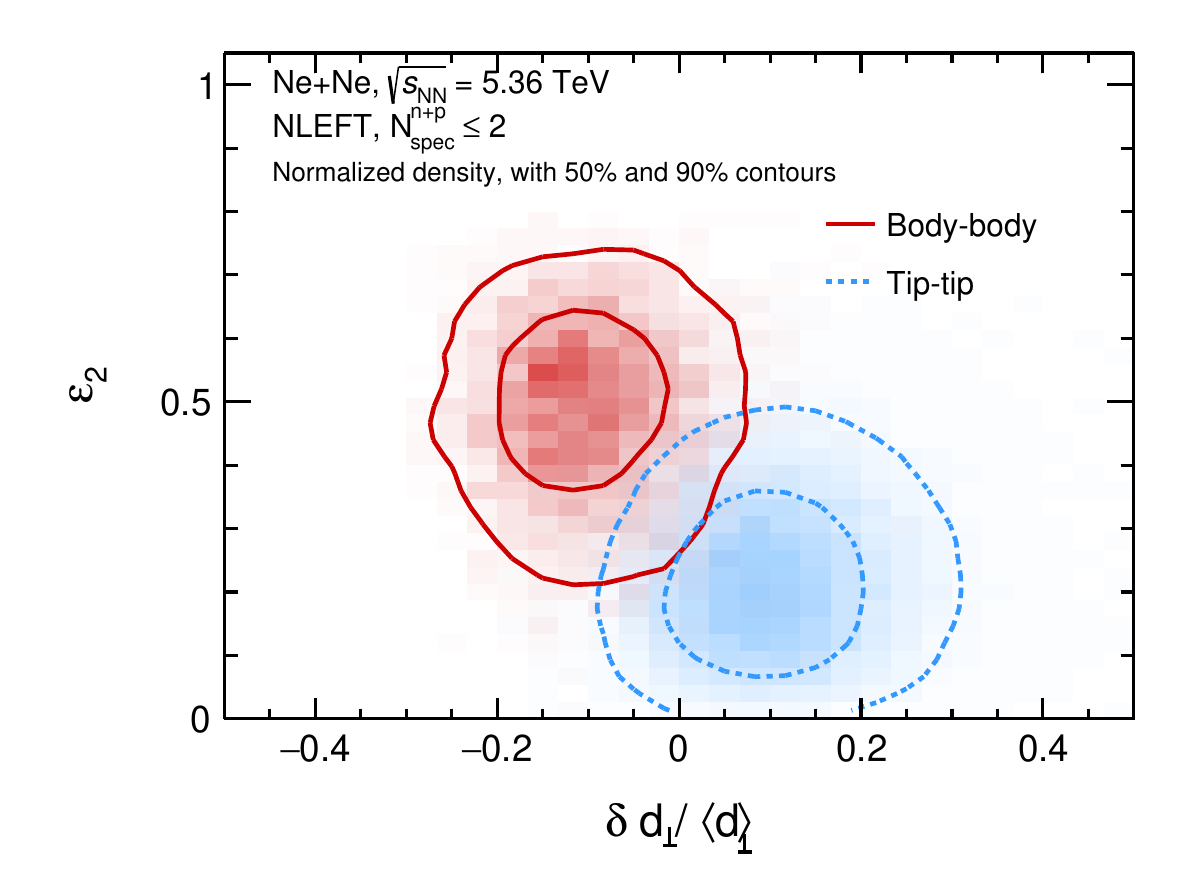}
  \includegraphics[scale=0.4]{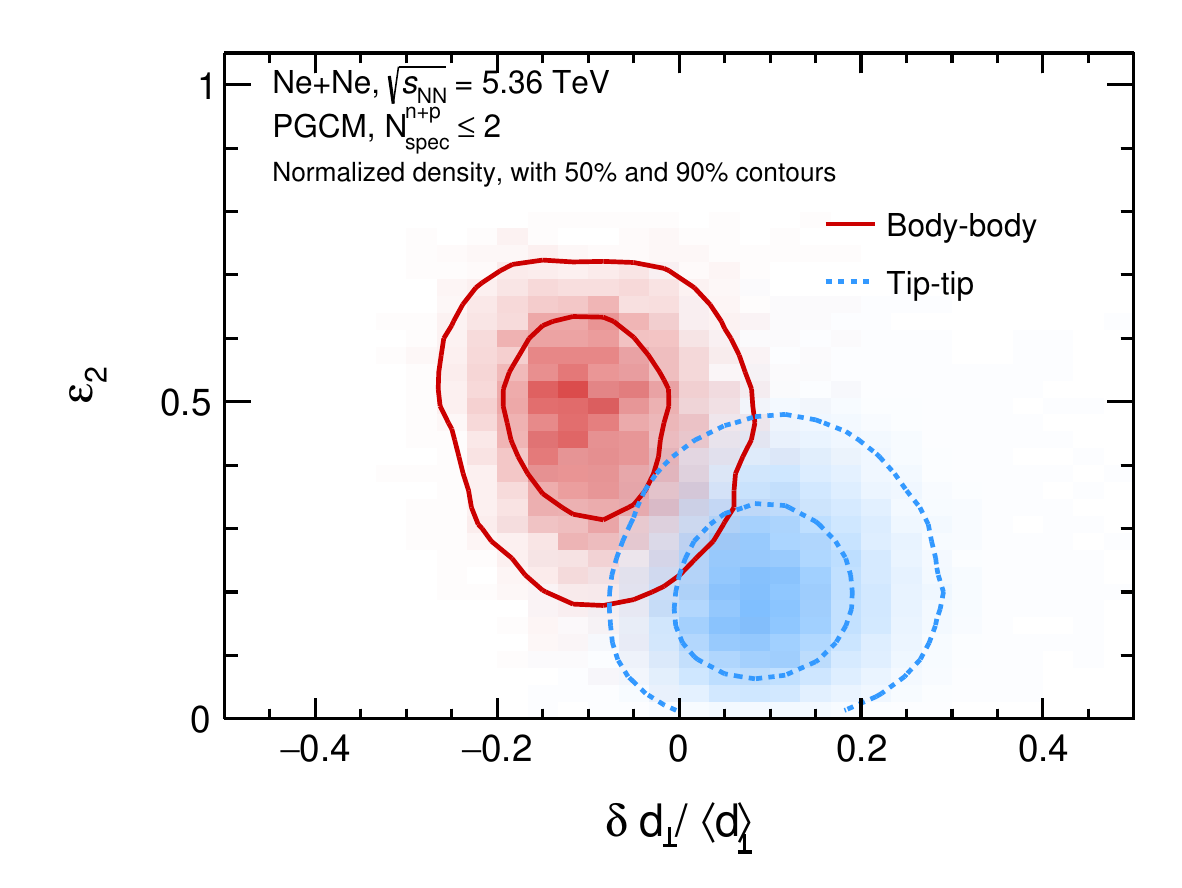}
  \caption{Density distributions of body--body and tip--tip events calculated with NLEFT (left) and PGCM (right) nuclear structure inputs in the
  $\varepsilon_2$--$[d_\perp]$ plane for $N_\mathrm{spec} \leq 2$.}
  \label{fig:ecc2_dperp}
\end{figure*}
\begin{figure*}[t!]
  \centering
  \includegraphics[scale=0.4]{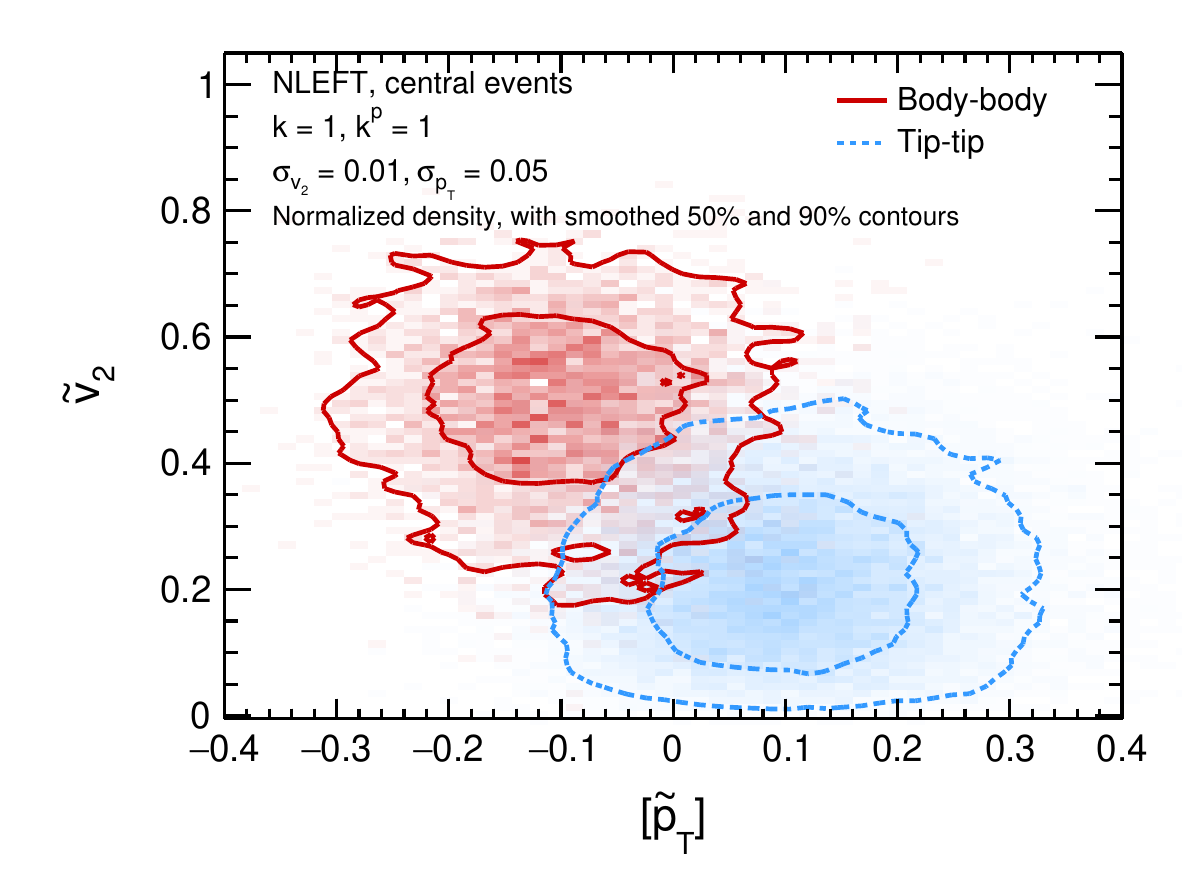}
  \includegraphics[scale=0.4]{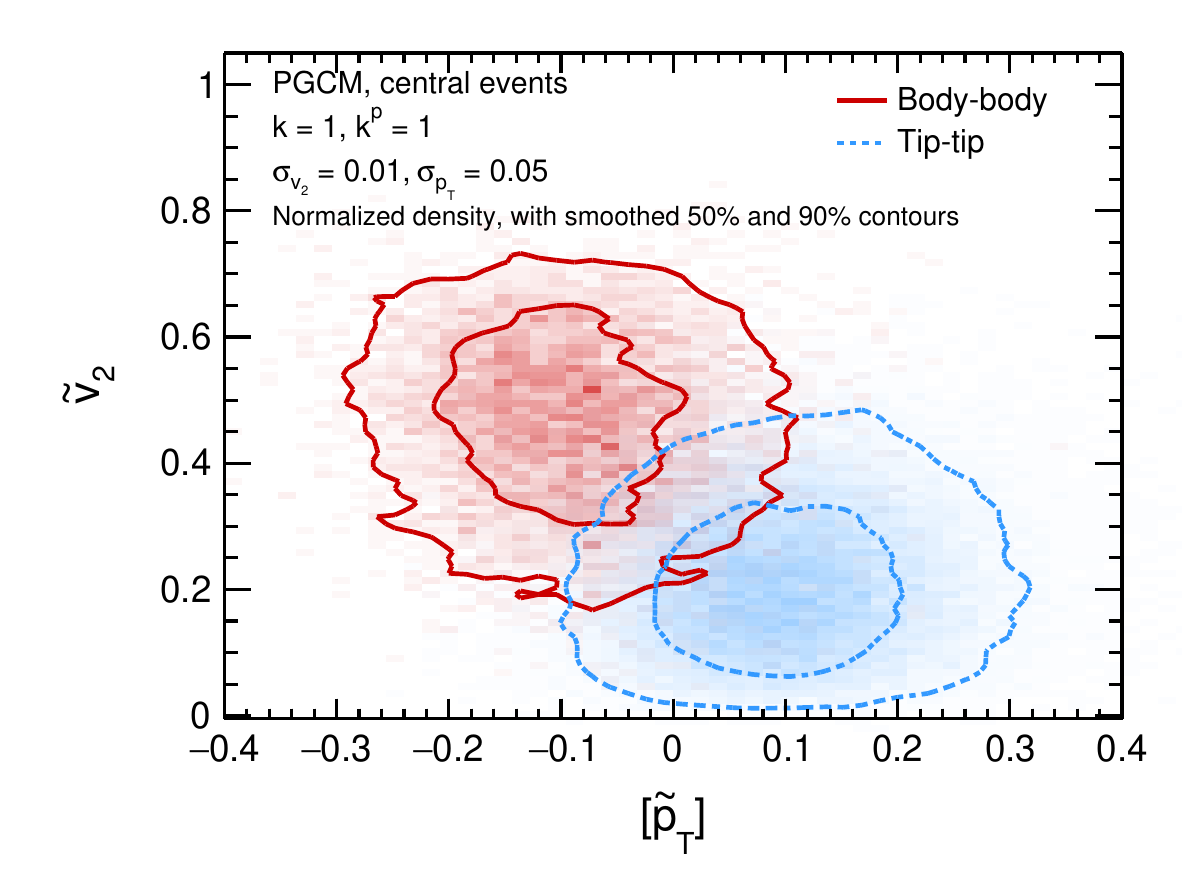}
  \caption{Density distributions of body--body and tip--tip events calculated with NLEFT (left) and PGCM (right) nuclear structure inputs in the
  $\tilde{v}_2$--$[\tilde{p}_\mathrm{T}]$ plane for $N_\mathrm{spec} \leq 2$. The $\sigma_{\tilde{v}_2}$ describes the width of the fluctuations used to smear $\varepsilon_2$ and $\sigma_{\tilde{p}_\mathrm{T}}$ is the width used to smear $[d_\perp]$.}
  \label{fig:v2_meanpt}
\end{figure*}

The combined dependence on $\varepsilon_2$ and $[d_\perp]$ is shown in
Fig.~\ref{fig:ecc2_dperp}. Body--body events are concentrated towards large
$\varepsilon_2$ and smaller $[d_\perp]$, whereas tip--tip events preferentially
occupy the region with small $\varepsilon_2$ and large $[d_\perp]$. The two
classes remain partially overlapping, but their different correlations produce a
visible separation in the two-dimensional plane.

This observation suggests two complementary selection strategies. A
body--body-enriched sample can be obtained by requiring large $\varepsilon_2$
together with small $[d_\perp]$, while the opposite selection enhances the
tip--tip contribution. The approximately linear separation visible in the figure
also motivates the Fisher discriminant introduced in
Sec.~\ref{sec:separators}.

\subsection{Final–state sensitivity}
To investigate whether the sensitivity to tip--tip and body--body
configurations survives the transition from the initial geometry to
final-state observables, a simple response model is used. This is done to assess how fluctuations in the relation between the initial and final states affect the selections. Motivated by the approximate response of elliptic flow to the initial eccentricity, the
toy elliptic-flow observable is defined as
\begin{equation}
  \tilde{v}_2 = k\varepsilon_2 + \delta v_2,
\end{equation}
where $k$ is a response coefficient and $\delta v_2$ represents
event-by-event fluctuations around the linear response. Similarly,
a toy observable for the event-wise mean transverse momentum is
constructed from the initial transverse compactness,
\begin{equation}
  [\tilde{p}_{\mathrm{T}}]
  = k'[d_\perp] + \delta p_{\mathrm{T}},
\end{equation}
where $k'$ controls the response to compactness and
$\delta p_{\mathrm{T}}$ describes the corresponding fluctuations.

For this study, $k=k'=1$, so the toy observables are expressed in rescaled units rather than calibrated to physical values of elliptic flow or mean transverse momentum. The fluctuation terms are taken to
follow Gaussian distributions with zero mean. Their widths control the strength of the smearing and can be varied separately to investigate the effect of a less precise response in either observable, here the relative fluctuations are $5$ times larger in the mean transverse momentum compared to elliptic flow. Increasing the smearing broadens the overlap between orientation classes and allows the stability of the selections to be examined.

This construction provides a simple test of the sensitivity to fluctuations in the mapping from initial geometry to final-state observables. It does not describe the full hydrodynamic evolution,
which can introduce non-linear response, correlated fluctuations, and a dependence on collision centrality~\cite{Gardim:2014tya,Noronha-Hostler:2015dbi}.

\begin{figure*}[t!]
    \centering
    \includegraphics[width=1\linewidth]{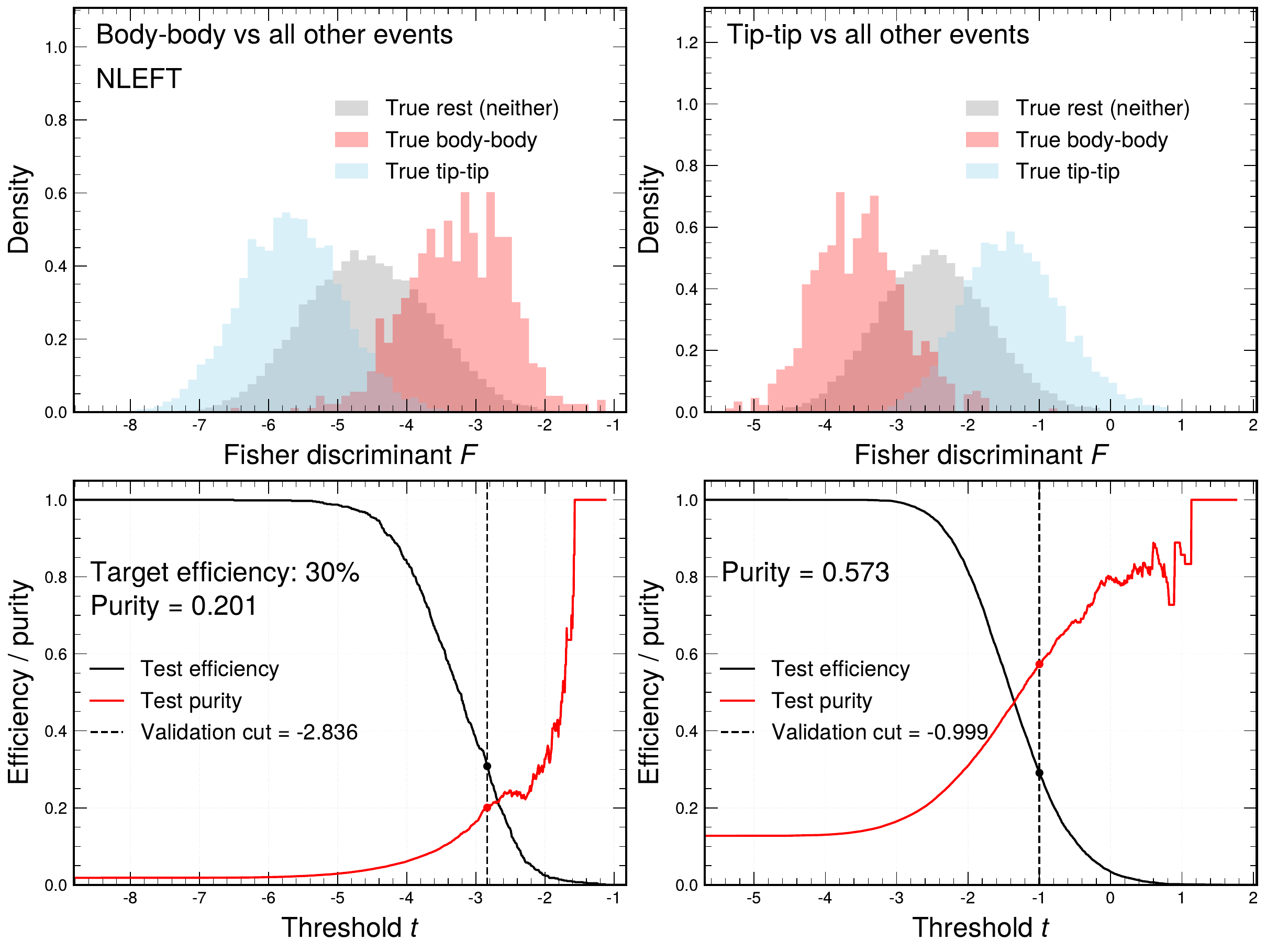}
    \caption{Fisher discriminant distributions and threshold scans for body--body (left) and tip--tip (right) selection (NLEFT). The upper panels show the distributions for true body--body, tip--tip, and remaining events, each normalized to unit area. The lower panels show the efficiency and purity for the selection $F \geq t$. Dashed vertical lines indicate thresholds chosen on the validation sample to achieve a target efficiency of $30\%$. All distributions and performance measures shown are evaluated on the independent test sample. Each discriminant is trained against all non-target events, including the opposite orientation class.}
    \label{fig:fisher_discriminant}
\end{figure*}

\section{Linear event separation and variable scans}
\label{sec:separators} Two independent Fisher linear discriminants are constructed to separate body--body and tip--tip events, respectively, from all other events. In each case, the background includes the
opposite orientation class and events belonging to neither class. The input observables are standardized using the training sample.

For target class $\alpha$, the discriminant is
\begin{equation}
F_\alpha(\boldsymbol{z})=\boldsymbol{w}_\alpha^T\boldsymbol{z}+c_\alpha,
\qquad
\boldsymbol{w}_\alpha \propto
(S_\alpha+S_{\beta})^{-1}
(\boldsymbol{\mu}_\alpha-\boldsymbol{\mu}_{\beta}),
\end{equation}
where $\boldsymbol{\mu}$ and $S$ denote the class means and covariance matrices in standardized coordinates. $\beta$ contains the rest within the centrality selection. The background covariance is calculated from the pooled non-target sample.

Events are selected by requiring $F_\alpha \geq t_\alpha$. Each threshold is
chosen on the validation sample to obtain a target efficiency of 30\%, and the resulting efficiency and purity are evaluated on an independent test sample. The corresponding purity is calculated as
\begin{equation}
P(t)
=
\frac{N_{\mathrm{target}}(t)}
{N_{\mathrm{selected}}(t)},
\end{equation}
where $N_{\mathrm{target}}(t)$ is the number of selected events belonging to the target orientation and $N_{\mathrm{selected}}(t)$ is the total number of selected events. These scans quantify the separation that can be obtained from a single
observable.

Figure~\ref{fig:fisher_discriminant} shows on the upper (right) left panel, the distributions of selecting (tip--tip) body--body vs the Fisher discriminant using the NLEFT data sample. The lower panels of the figure show the efficiency and purity curve with the optimized cut at $30\%$ efficiency. The purity of selecting body--body is at a modest $0.201$ and the purity of selecting tip--tip events is $0.573$. When comparing the separation of body--body and tip--tip from one another, their overlap is quite small. The selection cut imposed in this study provides a rather hard cut on these distributions and imposes a serious bias on geometries that approach either the selected body--body or tip--tip. Therefore, the events of the true rest that lie close to either configuration could, in principle, be geometries that approach the desired shape, so that the selection might be better than indicated by the purity

The Fisher discriminant provides a transparent baseline because the contribution from each observable can be inspected directly. Its limitation is that it describes the separation using a single linear boundary, while the distributions in Fig.~\ref{fig:ecc2_dperp} contain overlapping and potentially non-linear structures. A machine learning approach was tested, but did not provide a significant increase in purity.


\section{Conclusion}
\label{sec:conclusion}

In this work, the geometry of $^{20}$Ne--$^{20}$Ne collisions was studied using a Monte--Carlo Glauber model with PGCM and NLEFT nuclear configurations. The main purpose was to investigate whether the orientation and deformation of the colliding nuclei can be accessed through selections in the initial eccentricity $\varepsilon_2$ and the transverse compactness $d_\perp$.

The body--body and tip--tip configurations produce distinct, but overlapping, distributions. Body--body events are found at large $\varepsilon_2$, while tip--tip events are enhanced at large $[d_\perp]$. The combined $\varepsilon_2$--$[d_\perp]$ distribution provides a clearer separation than either observable alone and shows why multidifferential selections are
required. Furthermore, the comparison of spherical and deformed nuclear configurations demonstrates that sensitivity to deformation is not limited to the most central collisions. In particular, selecting the highest-$\varepsilon_2$ events preserves a visible deformation effect down to approximately $N_\mathrm{part}\sim13$.

A selection was constructed using a Fisher linear discriminant. The strongest linear separation was obtained by combining $[d_\perp]$ and $\varepsilon_2$, yielding a purity when constrained by requiring an efficiency of $30\%$ of \nleftbodypurity for selecting body--body configurations yielding an enrichment of around $11$ times of the data, while a purity of \nlefttippurity was obtained for the tip--tip configuration, which enriches the tip--tip data by a factor of $4.5$. This confirms that these observables contain complementary information about the collision geometry.

Finally, these results show that the orientation of deformed nuclei leaves measurable information in the initial collision geometry, even after accounting for event-by-event fluctuations and hydrodynamic smearing.

\bibliography{bibliography}
\clearpage
\appendix
\onecolumngrid
\section*{Supplemental Material}
\begin{figure*}[htb!]
    \centering
    \includegraphics[width=1\linewidth]{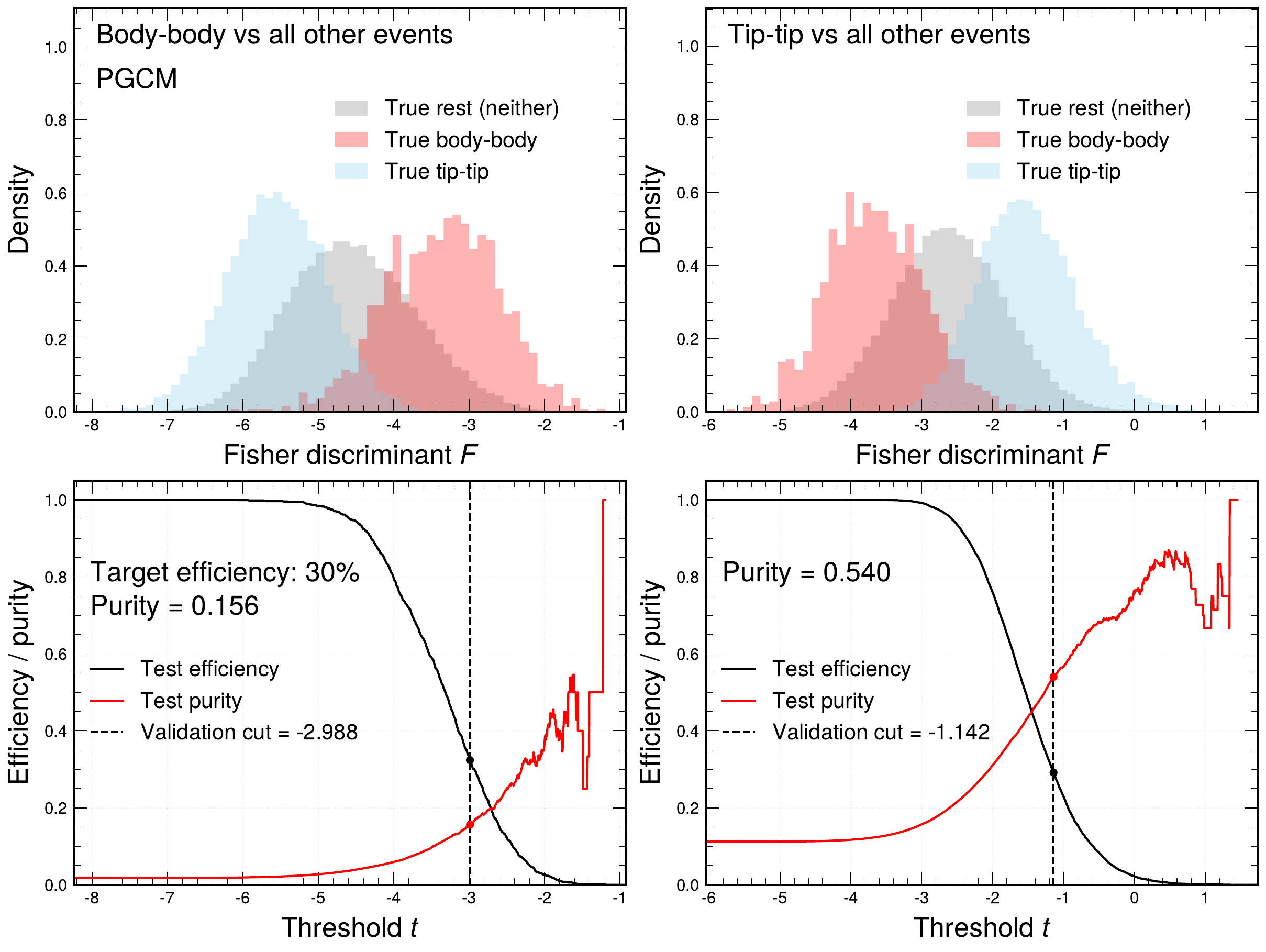}
    \caption{Fisher discriminant distributions and threshold scans for body--body (left) and tip--tip (right) selection (PGCM). The upper panels show the distributions for true body--body, tip--tip, and remaining events, each normalized to unit area. The lower panels show the efficiency and purity for the selection $F \geq t$. Dashed vertical lines indicate thresholds chosen on the validation sample to achieve a target efficiency of $30\%$. All distributions and performance measures shown are evaluated on the independent test sample. Each discriminant is trained against all non-target events, including the opposite orientation class.}
    \label{fig:fisher_discriminant_smeared}
\end{figure*}

\end{document}